\documentstyle[aps,prc,preprint,tighten]{revtex}
\begin{document}
\preprint{KSUCNR-016-94 (revised)}
\draft
\title{Thermal photon production in
high-energy nuclear collisions}
\author{John J. Neumann,\thanks{Electronic mail (internet):
neumann@scorpio.kent.edu.} David Seibert,\thanks{Current address:
Department of Physics, McGill University, 3600 University St.,
Montreal, QC, H3A 2T8, Canada.  Electronic mail (internet):
seibert@hep.physics.mcgill.ca.} and George Fai\thanks{Electronic mail
(internet): fai@ksuvxd.kent.edu.}}
\address{Center for Nuclear Research\\ Department of Physics, Kent State
University, Kent, OH 44242}
\maketitle
\begin{abstract}
We use a boost-invariant one-dimensional (cylindrically symmetric)
fluid dynamics code to calculate thermal photon production in the
central rapidity region of S+Au and Pb+Pb collisions at SPS energy
($\sqrt{s}=20$ GeV/nucleon).  We assume that the hot matter is in
thermal equilibrium throughout the expansion, but consider deviations
from chemical equilibrium in the high temperature (deconfined) phase.
We use equations of state with a first-order phase
transition between a massless pion gas and quark gluon
 plasma, with transition temperatures in the range $150
\leq T_c \leq 200$ MeV.
\end{abstract}

\section{Introduction}

Electromagnetic probes (photons and leptons) have long mean free paths
in hadronic matter and are therefore well suited for studying the
earliest stages of ultrarelativistic nuclear collisions.  Preliminary
single photon spectra from 200 GeV/nucleon S+Au collisions at CERN's
Super Proton Synchrotron (SPS) have recently become available from the
WA80 experiment,\cite{wa80} and results from Pb+Pb collisions should be
available in the near future.  In the present work we calculate thermal
photon spectra in the central rapidity region for SPS collisions,
based on a boost-invariant fluid-dynamical description of the hot
matter; a similar approach was used in Ref.~\cite{dinesh}.

Earlier predictions of photon spectra\cite{phth} and of dileptons from
several resonances\cite{rho,phi,phi2}
show a useful sensitivity to the
assumed QCD transition temperature when the equation of state
incorporates a strong first-order phase transition
and transverse expansion is neglected. The correlation between the
apparent temperatures of photon spectra and the assumed transition
temperature was suggested as a thermometer
for the transition temperature.\cite{phth}
One question we wish to study here is whether a useful correlation
can still be found if transverse expansion is taken into account.

In our calculation we assume a boost-invariant longitudinal
expansion as discussed by Bjorken,\cite{Bj} coupled to a cylindrically
symmetric transverse expansion. The one-dimensional fluid-dynamics
code we developed uses the Godunov method as described by Blaizot and
Ollitrault.\cite{bo}  We assume thermal equilibrium throughout the
evolution of the system, but consider deviations from chemical
equilibrium in the high-temperature phase by allowing the quark and
antiquark densities to be a (fixed) fraction of the equilibrium
value. In the photon production rate we include the
effect of the $a_1$ resonance.
We calculate the transverse momentum distribution for photons
in the range $1 \leq p_T \leq 2$ GeV.
We investigate the sensitivity to different assumptions about
the initial temperature, freezeout temperature, and
quark fraction, and compare production rates to the preliminary
WA80 data.
We use standard high-energy conventions, $c=\hbar=k_B=1$.

\section{Fluid-dynamical evolution}

For high collision energy we expect longitudinal
boost invariance,\cite{Bj}
so the behavior
of the produced matter at different rapidities is the same
in the longitudinally comoving frame
for fixed proper time $\tau=\sqrt{t^2-z^2}$, where
$z$ is the distance along the beam axis.
At  $\tau=0$ the colliding nuclei reach the point
of maximum overlap and are assumed to form a
longitudinally expanding pancake.
The hot matter has thermalized at
$\tau=\tau_0$ ($\approx$ 0.2 fm/$c$\cite{shur,gkce}),
when transverse expansion begins, coupled to the
 longitudinal expansion.
We treat the hot matter as one-dimensional,
 assuming cylindrical symmetry and boost invariance
 with  equations of motion \cite{bo}
\begin{mathletters}
\label{eqomot}
\begin{eqnarray}
\frac{\partial}{\partial \tau} (r\tau T^{00})+
 \frac{\partial}{\partial r} (r\tau T^{r0}) &=&-rP, \label{mlett:1}
\\
\nonumber
\\
\frac{\partial}{\partial \tau} (r\tau T^{0r})
+ \frac{\partial}{\partial r}(r\tau T^{rr})
&=&\,\,\,\,\tau P. \label{mlett:2}
\end{eqnarray}
\end{mathletters}
\\
Here $P$ is the pressure, $r$ is the radial coordinate,
and the energy-momentum
tensor $T^{\mu\nu}$ is $T^{\mu\nu}=(e+P)u^\mu u^\nu -
P\,g^{\mu \nu}$,
with $e$ being the energy density,
$u$ the four-velocity,
and metric tensor $g^{\mu\nu}=$ diag$(1,-1,-1,-1)$ .
The  $r-$weighting takes into account the radial geometry, and the
$\tau -$dependence is due to the coupling between the longitudinal and
transverse expansions.

The fluid dynamics code uses the Godunov method as described by
Blaizot and Ollitrault.\cite{bo} The system is described by
the relativistically covariant energy-momentum tensor elements $T^{00}$
and $T^{0r}$.  The program calculates the changes in the cell averages
of these tensor elements, respectively $\Theta^{00}$ and $\Theta^{0r}$,
by calculating the flows $T^{0r}$ and $T^{rr}$ across the cell walls.
A relationship between the flows at the walls and the changes in
$\Theta^{\mu\nu}$ is found by integrating (\ref{eqomot}) over each
space-time cell, assuming plane similarity forms for the flow patterns.
We approximate the space-time averaged pressure $\langle P \rangle$,
by using the initial value of $T^{\mu\nu}$ to estimate
$\Theta^{\mu\nu}$ at intermediate times, and constructing $\langle P
\rangle$ from these estimates for $\Theta^{00}$ and $\Theta^{0r}$.
This estimated value of $\langle P \rangle$ is then used to solve for
$\Theta^{\mu\nu}$ at the end of the time step.  At the beginning of the
next time step, we set $T^{00}$ and $T^{0r}$ throughout each cell equal to
$\Theta^{00}$ and $\Theta^{0r}$ for that cell at the end of the previous
step.

{}From $\tau=0$ until the transverse expansion starts at $\tau=\tau_0$,
we assume a boost-invariant cylinder of radius $R_<$
(the radius of the smaller
nucleus), filled uniformly with QGP at temperature $T=T_0$.
This is approximately
compatible with the initial entropy density for short times.\cite{s0}
We determine $T_0$ by assuming entropy conservation
for $\tau > \tau_0$,
hence
\begin{equation}
s(T_0)=\frac{3.6 \, dN/dy}{\pi \, R_<^2 \, \tau_0},
\end{equation}
where $s$ is the entropy density, with total (charged plus neutral)
multiplicity density $dN/dy=240$ and 1600 for S+Au and
Pb+Pb collisions, respectively.

The equations of state (EOS's) that we use here are of the form
\begin{mathletters}\label{EOS}
\begin{eqnarray}
T<T_c:\quad&&\left\{ \begin{array}{lll}
e &=&\displaystyle \frac{\pi^2}{10} T^4,\\[12pt]
P &=&\displaystyle \frac{e}{3},
\end{array} \label{EOS:had} \right. \\[12pt]
T=T_c:\quad&&\left\{ \begin{array}{lll}
\displaystyle \frac{\pi^2}{10} T_c^4 &\leq e \leq&
  \displaystyle \frac{\pi^2}{30} g_q T_c^4 +B,\\[12pt]
\displaystyle \frac{\pi^2}{30} T_c^4 &=P=&
  \displaystyle \frac{\pi^2}{90} g_q T_c^4 -B,
\end{array} \label{EOS:coex} \right. \\[12pt]
T>T_c:\quad&&\left\{ \begin{array}{lll}
e &=&\displaystyle \frac{\pi^2}{30} g_q T^4 +B,\\[12pt]
P &=&\displaystyle \frac{\pi^2}{90} g_q T^4 -B,
\end{array} \label{EOS:qgp} \right.
\end{eqnarray}
\end{mathletters}
\\
where $g_q$ is the number of massless degrees of freedom
in the deconfined phase.
We treat only the case of zero baryon density, so the entropy density is
\begin{equation}
s = \frac {e+P} {T},
\end{equation}
independent of the phase of the matter.  Below $T_c$, the EOS is that of
a massless pion gas.  Because recent calculations have
predicted that the quarks may reach only a
fraction of their equilibrium number by the beginning of
transverse expansion\cite{shur,gkce}, we take
$g_q =16+21x$, where $x$ is a parameter
that we vary between 0 and 1 to simulate
 the effect of reducing the quark density in the QGP
 below the equilibrium value
($x$=1 is equilibrium for two flavors of massless quarks).
The vacuum energy density in the deconfined phase, $B$, is related to
the transition temperature, $T_c$, by requiring equal pressures in
the deconfined and hadronic phases at $T=T_c$:
\begin{equation}
B=\frac {\pi^2 (g_q-3)} {90} T_c^4. \label{TcB}
\end{equation}

For all the fluid dynamic calculations we have taken
the radial cell size $\delta r$ and
time step $\delta \tau$ small enough to
be within a few percent of the continuum limit.
We also required
 $\delta \tau \ll \delta r/2$,
 guaranteeing that a similarity pattern from one
 side of a cell does not
overlap  the opposite side of the cell or
 the pattern from the  other side.
 Also, because the radial pattern size is much smaller
than $\delta r$,
we may assume plane similarity solutions as a
good approximation of the intercell
flow in spite of the radial geometry of the system.

\section{Thermal photons} \label{stp}

The central region photon $p_T$ distribution from
 boost-invariant hot matter is
\begin{eqnarray}
\left. \frac {d^2N_{\gamma}} {p_T dp_T dy} \right|_{y=0}
&=& \int d\eta \int d\tau \, \tau \int dr \, 2\pi r
\int_0^{\pi} d\theta \, \sin\theta \int_0^{2\pi} d\phi
\int_0^{\infty} dp \, p^2 \, \frac {dR} {d^3p} \nonumber \\[12pt] &&
\delta\left( \frac 1 2 p_T^2 - \frac 1 2 p_T^{\prime 2} \right) \,
\delta\left[ \eta +\tanh^{-1} \left( \frac {\cos\theta}
{\gamma \left| 1 \, + \, v \sin\theta\cos\phi \right|} \right) \right],
\end{eqnarray}
where
\begin{equation}
p_T^{\prime 2} = p^2\gamma^2 \left[  \sin^2\theta +2v\sin\theta\cos\phi
+v^2 \left( 1 -\sin^2\theta\sin^2\phi \right)\right].
\end{equation}
Here $R$ is the photon production rate per unit four-volume in the CM
frame of the hot matter, and $v$ is the transverse velocity of the matter
in the cell characterized by proper time $\tau$, space-time rapidity $\eta$
and radial position $r$
[measured in the frame
moving with transverse velocity zero and longitudinal velocity
$\tanh\eta$ in the lab].  Evaluating the integrals over $\eta$ and $p$
with the $\delta$-functions, we obtain
\begin{eqnarray}
\left. \frac {d^2N_{\gamma}} {p_T dp_T dy} \right|_{y=0}
&=& \int d\tau \, \tau \int dr \, 2\pi r \int_0^{\pi}
d\theta \, \sin\theta \int_0^{2\pi} d\phi \, \frac {p_*^3} {p_T^2} \,
\left( \frac {dR} {d^3p}\right)_{p=p_*},  \label{prodint} \\[12pt]
p_* &=& \frac {p_T} {\gamma \left[ \sin^2\theta +2v\sin\theta\cos\phi
+v^2 \left( 1 -\sin^2\theta\sin^2\phi \right) \right]^{1/2}}.
\end{eqnarray}

The photon production rate from thermally and chemically equilibrated
quark-gluon plasma is \cite{kls}
\begin{equation}
\frac {dR} {d^3p} ~=~ \frac {5 \, \alpha \, \alpha_s} {18 \, \pi^2 \, p}
\, T^2 \, e^{-p/T} \, \ln \left( \frac {2.912 \, p} {g^2 \, T} \, + \,
1 \right)~, \label{ERse}
\end{equation}
where $\alpha \, = \, 1/137$, and $\alpha_s \, = g^2/4\pi \, = \, 0.4$.
[We use the semi-empirical formula of Ref. \cite{kls},
which is almost identical to the exact rate.]

 Equation (\ref{ERse}) is first order in $\alpha_s$ (or $g^2$), and
as $g$ is of order unity, its accuracy is of the order of unity.
We also use it for the hadron gas, as
its uncertainty is larger than
the difference between the first-order QGP and hadron gas
production rates.\cite{nkl}
The contribution of the $a_1$ resonance as an intermediate state
was discussed recently by several
authors.\cite{xiong,song,dinesh} Here we include the contribution of
the $a_1$ meson as parametrized in Ref. \cite{xiong}.

The calculation follows the space-time evolution of the collision
based on eq. (\ref{prodint}) and results in nearly-exponential spectra.
We then mimic the experimental procedure and extract a temperature
(the fit temperature, $T_{fit}$) from the obtained spectra.
For fitting purposes, we take (\ref{prodint}) with $v=0$
(i.e.\ ignoring transverse expansion),
resulting in the formula \cite{phth}
\begin{equation}\label{tfit}
\frac{d^2N}{p_T dp_T dy} \sim
\frac{T^{5/2}}{p_T^{1/2}}
e^{-p_T/T} \ln \left( \frac{2.912\,p_T}{g^2 T} + 1.12 \right),
\end{equation}
\\
which is accurate to about one percent for the range we consider here.

The photon production in QGP is dominated by quark-gluon collisions.
To simulate scenarios where the quark number is smaller than the
equilibrium value, we multiply the rate (\ref{ERse}) by $x$,
the fraction of the equilibrium number of quarks, for
the QGP component of the fluid.  This representation of the
deviations from chemical equilibrium is valid whenever the
quarks
and antiquarks are non-degenerate, so that Maxwell-Boltzmann statistics
can be used in place of Fermi statistics, and when the production
from quark-antiquark annihilation (which is proportional to $x^2$) is
small.

\section{Results} \label{sr}

Our standard calculation uses $\tau_0$= 0.2 fm/$c$, $x$=1, freezeout
temperature $T_{fo}=100$ MeV, and includes the $a_1$ resonance;
if a different value of one of the parameters is given,
it is to be assumed that the others are held at the standard values.
In Fig.~1, we vary $T_c$ from 150 to 200 MeV and calculate the
resulting values of the temperature, $T_{fit}$, needed in eq.\
(\ref{tfit}) to fit the resulting spectrum for a
central S+Au collision at SPS energy. Our standard value for the
equilibration time, $\tau_0$= 0.2 fm/$c$, implies an initial temperature
of $T_0=355$ MeV for the S+Au system.

In the absence of transverse expansion, $T_{fit}$ is a monotonically
increasing function of $T_c$, so one can infer $T_c$ given $T_{fit}$
from the measured photon spectrum.\cite{phth}  This can be seen in
the lowest curve, where we have run our simulation without transverse
expansion.

Including transverse expansion, we find that we can no longer infer
$T_c$ from $T_{fit}$, due to the non-monotonicity of the relation
(except for long thermalization times; see the curve with
$\tau_0$=1.0 fm/$c$, corresponding to $T_0=208$ MeV).
This demonstrates the large effect of the transverse motion of the fluid
on the photon spectrum: lowering $T_c$ decreases the bag constant, which
increases the ratio $P/e$ in the QGP, producing more transverse motion.
The increase in $p_T$ caused by the fluid motion is enough
to overcome the decrease of $T_{fit}$ with decreasing $T_c$ found
without transverse expansion.  This
non-monotonicity occurred for all parameter sets we
tried, although for some (e.g., $\tau_0=1$ fm/$c$)
the minimum appears outside the range of $T_c$ investigated
here. As the initial temperature is lowered, the value of $T_c$ below which
the effects of flow dominate also decreases.
This may be attributed to the fact that, as $T_0$
approaches $T_c$, less flow develops before the mixed phase is reached.

Contrary to experience without transverse expansion,\cite{phth}
increasing $T_{fo}$ to 150 MeV decreases $T_{fit}$,
another indication that the transverse motion has a large effect
on the spectrum.  Apparently the lower-temperature hadronic matter
expands outward at a velocity high enough that the
contribution to
$p_T$ from this motion more than makes up for the low
temperature.
Thus, removing the contribution from the late stages (by raising
$T_{fo}$) decreases $T_{fit}$.

For $x \neq 1$, the quantity we call  $T_c$ on the figure
is the chemical equilibrium
($x=1$) transition temperature,
\begin{equation}
T_c = \left( \frac {90} {34 \pi^2} B \right)^{1/4}, \label{Tcneq}
\end{equation}
rather than the larger non-equilibrium transition temperature used
in the simulation, which is given by eq.~(\ref{TcB}).  We do this
because the equilibrium transition temperature is more useful for
comparisons with theoretical calculations.  The curve for $x=0.9$ shows
that $T_{fit}$ is about 3 MeV higher than for $x=1$.  The increase
occurs because the temperature of the mixed phase, which makes a large
contribution to the photon spectrum, is raised when $x$ is lowered.  We
do not show results for other values of $x$; the change in $T_{fit}$ is
approximately linear in $x$.

The effect of increasing $\tau_0$ to 1 fm/$c$ is dramatic: $T_{fit}$
becomes
monotonic within the range of interest, and is lower than
for the
other parameter sets considered. As mentioned above, this is due to
the decrease in the role of
the transverse expansion, which is assumed to start at
$\tau_0$.

Figure 2 displays the effect of varying the parameters on the spectra,
keeping the value of the critical temperature fixed ($T_c=170$ MeV).
We plot $d^2 N_{\gamma}/p_Tdp_Tdy$ vs.\ $p_T$ for different scenarios.
The spectra are nearly exponential. It is seen that increasing the
equilibration time (i.e.\ decreasing the initial temperature) decreases
the high $p_T$ contribution to the photon spectrum and thus decreases the
fit temperature. Increasing the freezeout temperature removes a fraction of
the radiated photons and decreases the yield, also as expected.
Excluding the $a_1$ resonance decreases the yield by about one-third
in the transverse-momentum range investigated.
This is roughly consistent with several other
calculations.\cite{xiong,song,shur2,dinesh}

Figure 3 shows $d^2 N_{\gamma}/p_Tdp_Tdy$ vs.\ $p_T$ (per central
collision) for the preliminary WA80 data (assuming a minimum bias
cross-section of 3600 mb), along with calculated spectra
using selected parameter sets. The data
between 1 and 2 GeV are best fitted with $T_{fit}=213$ MeV, using
eq.~(\ref{tfit}). We use this value of $T_{fit}$ and Figure 1 to extract
the value of $T_c$ (between 150 and 200 MeV)
for each parameter set. The experimentally determined fit temperature,
$T_{fit}=213$ MeV is below the values predicted by the
model for some parameter choices.
In these cases $T_c$ was estimated by minimizing $T_{fit}$.

For the standard parameter set and for $x=0.9$ we used $T_c$=170 MeV.
Note, however, that for $x \neq 1$, the non-equilibrium transition temperature
is larger than this, as explained earlier. The difference is sufficient
to counteract the decrease in the number of degrees of freedom in the
plasma, leading to a somewhat higher yield than with the standard parameters.
Due to the large error bars of the preliminary WA80 data,
the calculated spectra are consistent with the data points, although
generally below them for these parameters.
In the absence
of transverse expansion, photon production is increased by approximately
a factor of two by including pion masses,\cite{phth}
so agreement with the preliminary WA80 data will probably be further improved
with a more realistic equation of state.

Increasing the equilibration time to $\tau_0$=1 fm/$c$ (and accordingly
decreasing the initial temperature) leads to
$T_c$=190 MeV. The increase in $T_c$ increases the photon yield.
In addition, the lower initial temperature and the higher critical
temperature results in less flow being built up by the time the system
reaches the mixed phase than with the standard parameter set. Therefore,
the system spends a longer time in the mixed phase for this set of
parameters. This also increases the yield, as pointed out by Shuryak and
Xiong.\cite{shur2} To quantify the difference between these scenarios,
we ``measured'' (in the code) the total four-volume associated with the
mixed phase
and found it to be about 25\% larger for $\tau_0$=1 fm/$c$ than for
$\tau_0$=0.2 fm/$c$ .
Our results with the long equilibration time ($\tau_0$=1 fm/$c$) tend to
overestimate the data. Excess photons come from both the QGP and the
hadronic phases in our calculation relative to Ref. \cite{shur2}.

Increasing the freezeout temperature to $T_{fo}$=150 MeV decreases the
photon yield partly because a contribution is removed, and partly because we
use $T_c=160$ MeV (rather than 170 MeV) in this case.
Note that 160 MeV$\leq T_c\leq 170$ MeV all give approximately the same
$T_{fit}$ for $T_{fo}$=150 MeV, though
the photon spectrum has a larger magnitude when $T_c$ is increased
due to the large contribution from the mixed phase.
Thus, even if $T_c$ cannot be
determined by $T_{fit}$ alone, it may be possible to ascertain $T_c$
by comparing to the total production rate as well.

Figure 4 displays predicted spectra for a central Pb+Pb
collision
at SPS energy, using the parameter sets from Fig.~3.
The initial temperatures
corresponding to $\tau_0$= 0.2 fm/$c$ and $\tau_0$= 1 fm/$c$ are
$T_0=441$ MeV and $T_0=$258 MeV, respectively.
The shapes of the
spectra and the relative production rates are similar to those for the S+Au
collisions.  For comparison, we also show the spectrum obtained in the
absence of transverse expansion, taking $T_c=200$ MeV.  We find that
transverse expansion has a large effect on the production
rate, in apparent disagreement with Ref.~\cite{dinesh}.

Our conclusion is that transverse expansion destroys the correlation
suggested in Ref.~\cite{phth}, so that $T_c$ cannot be determined just
from measurements of the shape of the single photon spectrum, as
parametrized by $T_{fit}$.  The monotonic dependence of $T_{fit}$ on
$T_c$ is destroyed because the transverse flow increases
as $T_c$ decreases.  However, the total production rate is also
sensitive to $T_c$, so it is possible that comparison to the production
rate instead of just the shape of the transverse momentum spectrum may
be sufficient to determine $T_c$ from data. In the transverse-momentum
range studied, the inclusion of the $a_1$ resonance increases the photon
yields by about 50\%. Preliminary WA80 data appear to rule out large
values for the equilibration time.

\acknowledgements

We thank T. Awes for useful discussions of the preliminary
WA80
results and J. Pol{\'{o}}nyi for helpful comments.  This work was
supported in part by the U.S. Department of
Energy under Grant No.\ DOE/DE-FG02-86ER-40251.

\newpage
\section*{Figure Captions}

\begin{description}
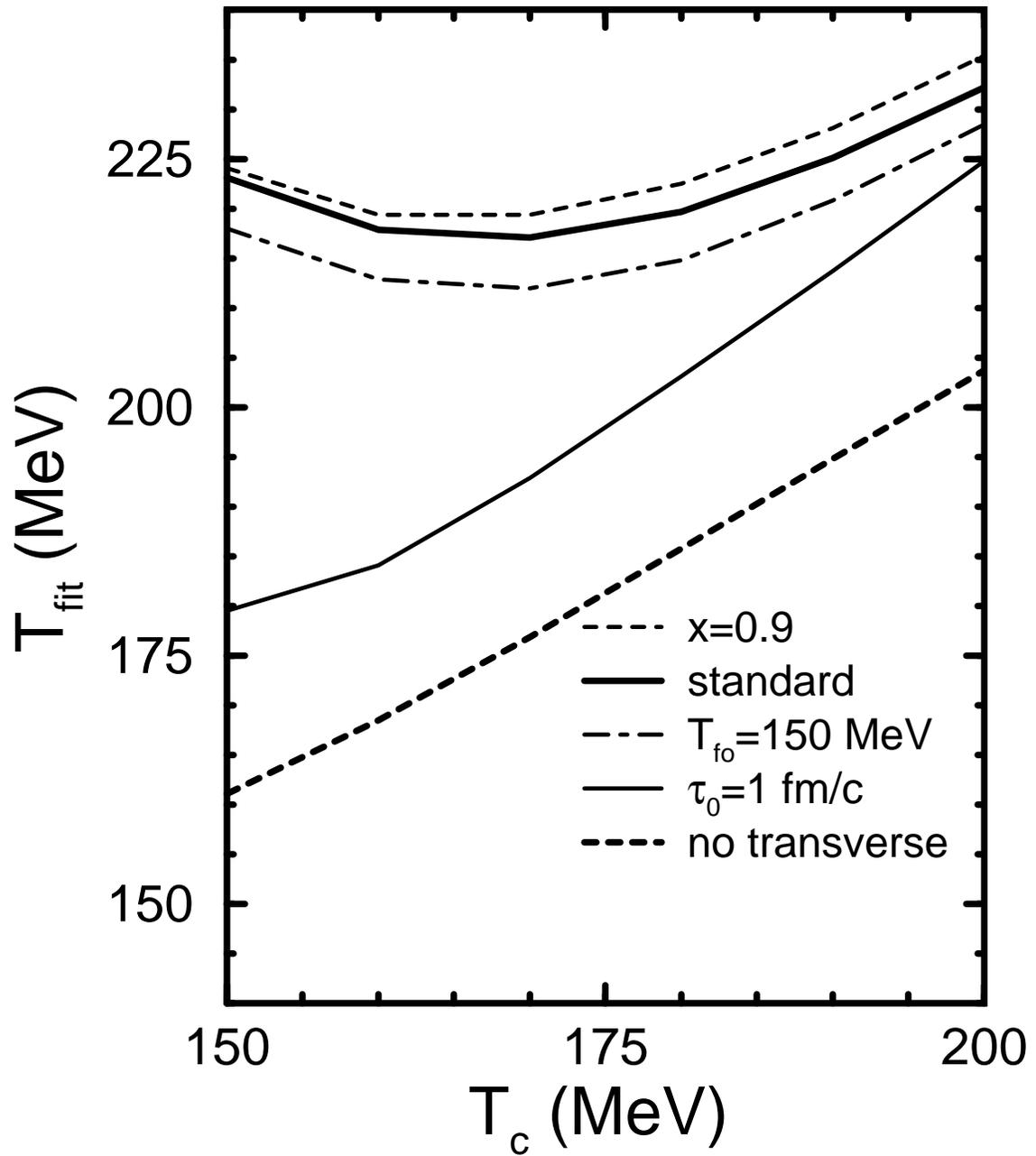
\item[Fig. 1:] $T_{fit}$ vs.\ $T_c$ for central S+Au collisions
at SPS energy with various parameter sets. The standard set consists of
$\tau_0$= 0.2 fm/$c$, (corresponding to $T_0=355$ MeV), $x$=1,
$T_{fo}=100$ MeV, and includes the $a_1$ resonance; the other sets differ from
the standard one in the value of the parameter indicated.
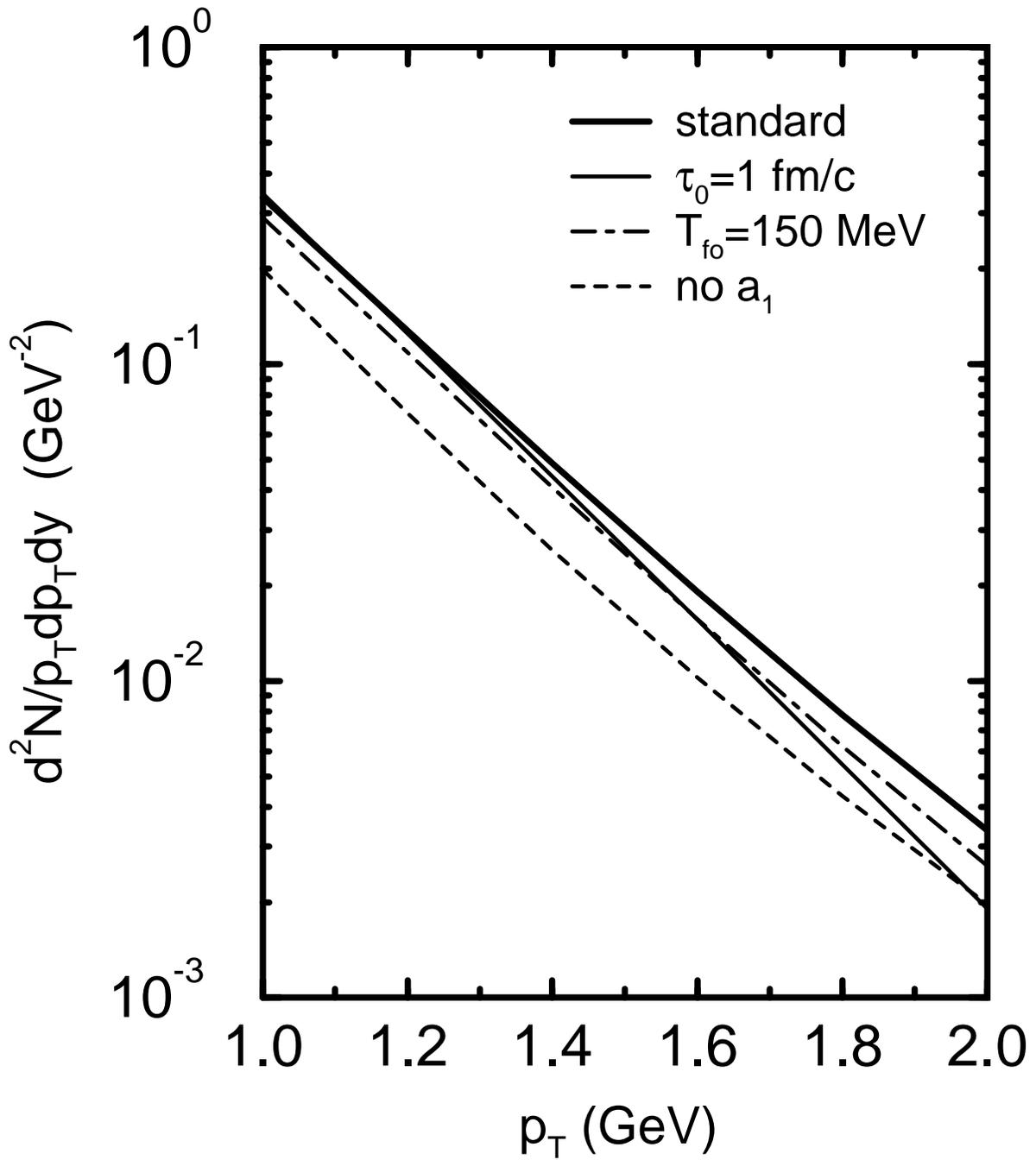
\item[Fig. 2:]  Photon spectra with different parameter sets keeping
$T_c=170$ MeV constant. The standard set is the same as in Fig. 1.
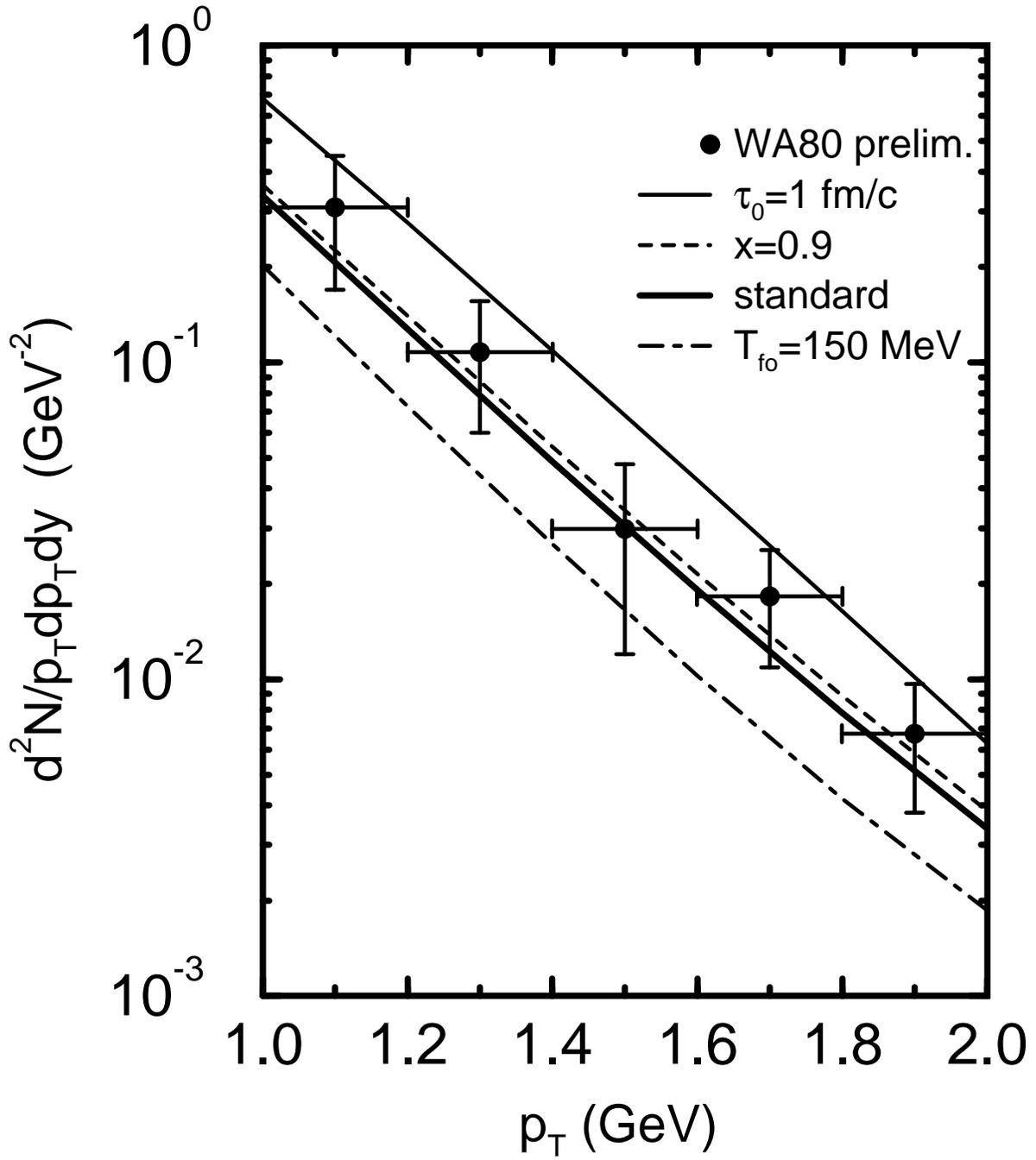
\item[Fig. 3:]  The preliminary WA80 data \cite{wa80}
and selected fits. For each curve the value of $T_c$ giving $T_{fit}=$213 MeV
from Fig. 1 is used as explained in the text.
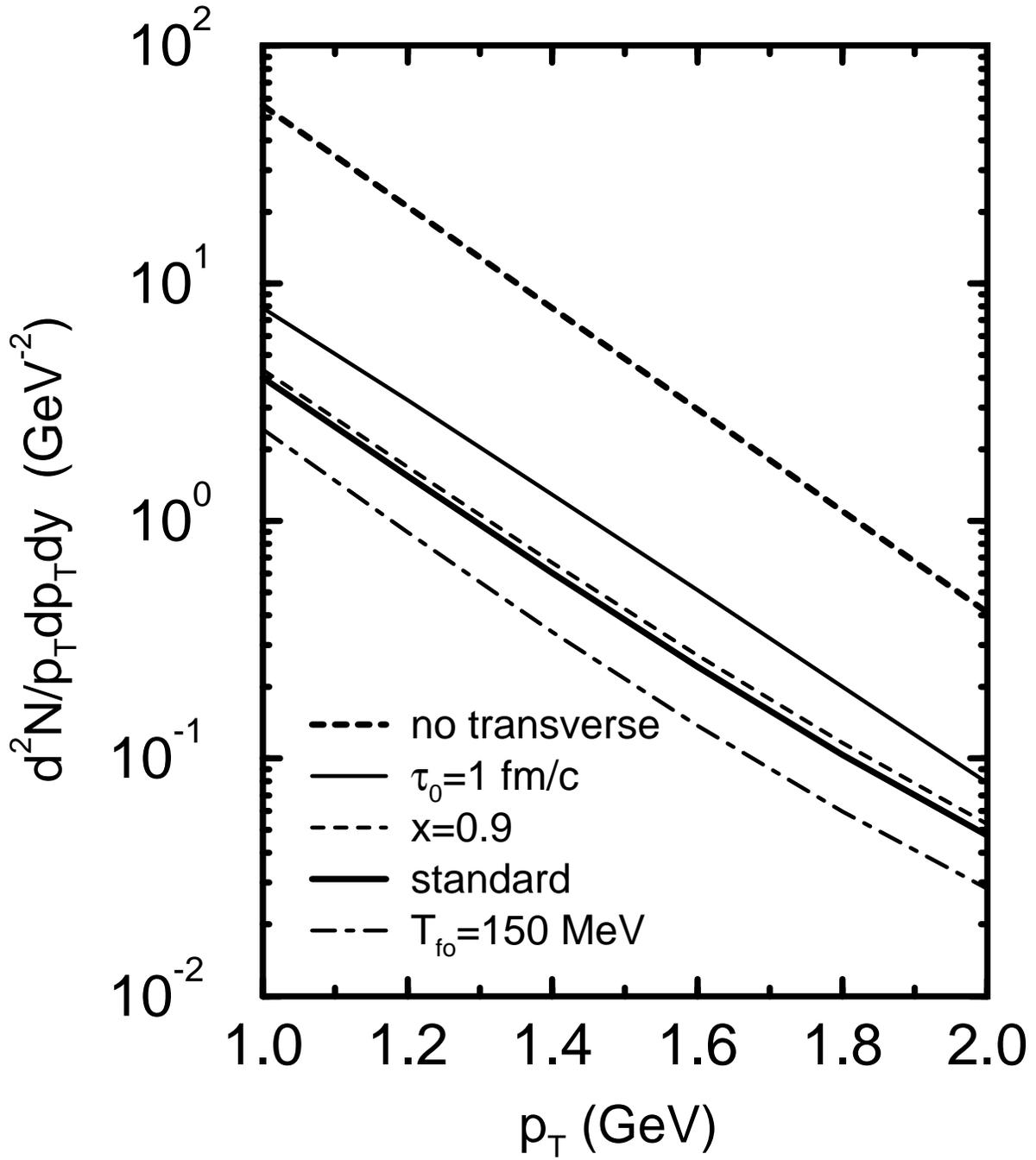
\item[Fig. 4:] Predicted photon spectra for central Pb+Pb
collisions at SPS energy.
\end{description}
\end{document}